\documentclass[twoside]{article}
\usepackage{lineno,hyperref,booktabs,epsfig,graphicx,amsmath,multirow,array,subfig,xcolor}
\usepackage{amstext}
\usepackage{multicol,headrule,ams2la}
\usepackage{graphics}
\usepackage{bm}                       
\usepackage{amssymb}                  
\usepackage{amsbsy} 
\graphicspath{{figs/}}      
\graphicspath{{figs/}}      

\def\oddhead{{\rm\hfill\protect\small
Deep Fault Diagnosis for Rotating Machinery with Scarce Labeled Samples 
\hfill }}
\def\evenhead{{\rm \hfill \protect\small\it Chinese Journal of
Electronics  \rm\hfill 2016}}

\def\RE{\par\vskip1.5\kh\par\centerline{\bf
References}\par\vskip0.5\kh\par}
\def\REF#1{\par\hangindent\parindent\indent\llap{#1\enspace}\ignorespaces}
\DeclareSymbolFont{lettersA}{U}{txmia}{m}{it}
\DeclareMathSymbol{\piup}{\mathord}{lettersA}{25}
\DeclareMathSymbol{\muup}{\mathord}{lettersA}{22}

\begin{document}
\newtheorem{theorem}{Theorem}
\newtheorem{proposition}{Proposition}
\newtheorem{remark}{Remark}
\newtheorem{proof}{Proof}
\def\footnoterule{\kern 1mm \hrule width 12cm \kern 2mm}
\abovedisplayskip=8.0pt plus 2.0pt minus 1.5pt
\belowdisplayskip=8.0pt plus 2.0pt minus 1.5pt
 \def\thefootnote{}
\TagsOnRight
\def\kh{\baselineskip}
\def\pmb#1{\boldsymbol{#1}}

\thispagestyle{empty}
\noindent Chinese Journal of Electronics

\noindent Vol.XX, No.X, Jan.\ 20XX

\vskip 2cm

\begin{center}\huge\bf
Deep Fault Diagnosis for Rotating Machinery with Scarce Labeled Samples$^*$
\end{center}\footnotetext{\footnotesize $^*$Manuscript Received May, 2019; Accepted June, 2019.\
This work is supported by the National key research and development
program (2016YFE0200900),the National Natural Science Foundation of China1
(No.61806064, No.61806062, No.61751304, No.61873077),Open Foundation of Key
Laboratory of Advanced Public Transportation Science, Ministry of Transport, PRC.
}

\vskip 4mm

\begin{center}\large\rm
ZHANG Jing$^{1}$$^{**}$, TIAN Jing$^{1}$$^{**}$, WEN Tao$^{2}$, YANG Xiaohui$^{3}$, RAO Yong$^{4}$ and XU Xiaobin$^{1}$  
\\\vskip1mm
\footnotetext{\footnotesize $^{**}$Co-first authors contributed equally.}
{\small(1.~\it School of Automation, Hangzhou Dianzi University, Hangzhou 310018, China
\rm)}

{\small(2.~\it School of Electronic and Information Engineering, Beijing Jiaotong University, Beijing 100044, China
\rm)}

{\small(3.~\it School of Mathematics and Statistics, Henan University, Kaifeng 475004, China
\rm)}

{\small(4.~\it Hofon Automation Co., Ltd, Hangzhou 311121, China
\rm)}\footnotetext{\footnotesize \copyright~2019 Chinese Institute
of Electronics. DOI:10.1049/cje.2019.0X.0XX}
\end{center}

\begin{multicols}{2}

{\footnotesize\bf Abstract --- {\footnotesize\bf
Early and accurately detecting faults in rotating machinery is crucial for operation safety of the modern manufacturing system. In this paper, we proposed a novel Deep fault diagnosis (DFD) method for rotating machinery with scarce labeled samples. DFD tackles the challenging problem by transferring knowledge from shallow models, which is based on the idea that shallow models trained with different hand-crafted features can reveal the latent prior knowledge and diagnostic expertise and have good generalization ability even with scarce labeled samples. DFD can be divided into three phases. First, a spectrogram of the raw vibration signal is calculated by applying a Short-time Fourier transform (STFT). From those spectrograms, discriminative time-frequency domain features can be extracted and used to form a feature pool. Then, several candidate Support vector machine (SVM) models are trained with different combinations of features in the feature pool with scarce labeled samples. By evaluating the pretrained SVM models on the validation set, the most discriminative features and best-performed SVM models can be selected, which are used to make predictions on the unlabeled samples. The predicted labels reserve the expert knowledge originally carried by the SVM model. They are combined together with the scarce fine labeled samples to form an Augmented training set (ATS). Finally, a novel 2D deep Convolutional neural network (CNN) model is trained on the ATS to learn more discriminative features and a better classifier. Experimental results on two fault diagnosis datasets demonstrate the effectiveness of the proposed DFD, which achieves better performance than SVM models and the vanilla deep CNN model trained on scarce labeled samples. Moreover, it is computationally efficient and is promising for real-time rotating machinery fault diagnosis. }

\bf Key words --- {\footnotesize\bf
Fault diagnosis, Deep CNN, Spectrograms, Knowledge-transferring.}}

\begin{center}{\large\bf I.\ Introduction
}\end{center}


Rolling bearings and other rotation elements are common components of machinery running in the modern manufacturing industry and transportation equipment,{\it etc.} Faults in rotating machinery will cause fatal breakdown of productive systems and lead to high maintenance costs and economic loss$^{[1-8]}$. Therefore, early and accurately detecting these faults is crucial for operation safety of the modern manufacturing system. Moreover, it can also benefit condition-based maintenance by carrying out appropriate maintenance strategies accordingly, and active fault-tolerant control to guarantee the reliability and safety of the controlled system$^{[3,5-7]}$. Due to the complex types of faults, noisy signals from onsite sensors, and scarcity of fault samples, early and accurate fault diagnosis is still very challenging and is an active research area.

Fault diagnosis methods can be summarized as the following three categories in terms of the input signal domains: time-domain$^{[1,2,9-12]}$, frequency-domain $^{[4,13,14]}$ and time-frequency based methods$^{[7,15-17]}$. Besides, according to the classifier models, they can be categorized into shallow model-based methods and deep learning-based methods. The shallow classifier models include Support vector machines (SVM)$^{[18-20]}$, Boosting$^{[11,21,22]}$ and Extreme learning machine (ELM)$^{[17,23]}$, {\it etc.}, which can be treated as Artificial neural network (ANN) model with one hidden layer. The deep learning-based methods are built in the multi-layer deep neural networks with cascade structures, such as Deep belief nets$^{[24,25]}$, and Deep Convolutional neural network (CNN) $^{[26-29]}$. Classical methods usually consist of two parts: hand-crafted feature extraction and shallow classifier training. For example, Zhang {\it et al.} propose a bearing fault diagnosis model based on SVM classifier and the permutation entropy features after decomposing the input vibration signal into a set of intrinsic mode functions by using ensemble empirical mode decomposition$^{[19]}$. Tian {\it et al.} propose a rolling bearing fault diagnosis based on extreme learning machine$^{[17]}$. They utilize a self-adaptive time-frequency analysis method named local mean decomposition to decompose the vibration signals into a series of product functions to reveal instantaneous frequencies with physical significance. Then, features are extracted by applying singular value decomposition to the product functions. Liu {\it et al.} propose a diesel engine fault diagnosis method using intrinsic time-scale decomposition and multistage Adaboost relevance vector machine$^{[11]}$. First, the vibration signal of diesel engine is decomposed by the intrinsic time-scale decomposition method. Then, both time-domain and frequency-domain features are extracted and used as the input of a multistage Adaboost relevance vector machine model.

These methods share a common characteristic that handcrafted fault features should be extracted first before inputting and training the classifier model. The feature extraction process highly relies on prior knowledge and diagnostic expertise. It is probably implemented case by case that the feature representation designed for a specific fault diagnosis case may not generalize well to other cases. Besides, whether the selected classifier has the best modeling capacity of the designed features is unclear since they are not jointly optimized.

In the past few years, deep learning-based methods have achieved remarkable success in many research fields such as speech recognition$^{[30]}$, image classification$^{[31]}$ and object detection$^{[32]}$. To overcome the aforementioned issues, different deep neural network based method has been proposed for fault diagnosis$^{[1-3,13,33-36]}$. For example, Chen {\it et al.} propose to use convolutional neural networks for gearbox fault identification and classification, where both time domain and frequency domain features vectors are formed and used as the network input$^{[13]}$. Ince {\it et al.} propose to utilize 1-D convolutional neural networks for real-time motor fault detection $^{[2]}$. The method can learn feature representation from the raw data and be jointly optimized together with the classifiers. Experimental results demonstrate its effectiveness and high computational efficiency. Xia {\it et al.} propose a deep CNN method which learns discriminative feature from the temporal and spatial information of the raw data from multiple sensors$^{[34]}$. Experimental results on datasets from two types of typical rotating machinery, roller bearings and gearboxes convince its effectiveness.

Though the deep learning-based method can achieve highly competitive performances, they consume large number of training samples with fine annotations to train the network parameters. However, fault samples are scarce and hard to collect in the industry site. Besides, samples should be annotated carefully by experienced experts. Some early work of deep neural network utilizes unlabeled data to pretrain a deep autoencoder in a layer-wise manner, and then fine-tune it with labeled data. For example, Jia {\it et al.} pretrain a deep neural network consists of cascaded auto-encoders in an unsupervised manner, then fine-tune it supervised by the labeled data to enable discriminative feature learning and classifying abilities for rotating machinery faults$^{[33]}$. Sun {\it et al.} utilize sparse auto-encoder to learn features and then use them to train a neural network classifier for motor faults diagnosis$^{[1]}$. However, features learned by the pretrained network from massive unlabeled data may be less discriminative since it is hard to reveal the intrinsic characteristics of fault samples when they are so scarce and overwhelmed by the normal ones.

Shallow models such as SVM, ELM, {\it etc.}, require less labeled samples than deep learning models to train the parameters and exhibit good generalization abilities. Besides, handcrafted features used as the input of those shallow models are designed according to the prior knowledge and diagnostic expertise, which reveal some intrinsic characteristics of fault samples. Inspired by the observations, We extend the previous work$^{[8]}$ and propose a novel deep fault diagnosis (DFD) method for rotating machinery with scarce labeled samples by transferring knowledge from shallow models. First, spectrogram of the raw vibration signal is calculated by applying a Short-time Fourier transform (STFT). From those spectrograms, discriminative time-frequency domain features can be extracted and used to form a feature pool. Then, several candidate SVM models are trained with different combinations of features in the feature pool on scarce labeled samples. By evaluating the candidate SVM models on the validation set, the most discriminative features and best-performed SVM models can be selected, which are used to make predictions on the unlabeled samples. The predicted labels reserve the expert knowledge originally carried by the SVM model. They are combined together with the scarce fine labeled samples to form an augmented training set (ATS). Finally, A novel 2D deep CNN model is trained on the ATS to learn more discriminative features and a better classifier. The proposed DFD follows a complete data-driven manner and demonstrates its effectiveness on two fault diagnosis datasets. In addition, it is computationally efficient and is promising for real-time rotating machinery fault diagnosis.

In contrast to Ref.[8], we make several novel contributions including: 1) A comprehensive knowledge-transferring based deep fault diagnosis model is proposed which consists of constructing feature candidate pool, pre-training shallow models and knowledge-transferring. Each part is elaborated with more detailed description and necessary mathematics. For example, in contrast to Ref.[8] which only utilizes integral features, we follow Ref.[34] and use 15 kinds of features to construct the feature pool and select the corresponding pretrained models. 2) Thorough experiments on two fault diagnosis benchmark datasets including the roller bearing condition dataset from Case Western Reserve University (CRWU) has been conducted to validate the proposed method, which is more convincing than Ref.[8]. 3) More detailed analysis and discussions were presented to explain the merit and demerit of the proposed method.

The rest of the paper is organized as follows: Section II presents the related work including hand-crafted features, short-time Fourier transform and convolutional neural network for fault diagnosis. The proposed DFD method based on knowledge-transferring is presented in Section III. In Section IV, we present the experimental results of DFD on two datasets. Finally, we conclude this paper in Section V and indicate several potential directions for future research.

\begin{center}{\large\bf II.\ Related Work
}\end{center}

{\bf 1.\ Hand-crafted features and shallow models}

Feature extraction plays a particularly important role in fault diagnosis. Discriminative hand-crafted features are able to greatly improve the accuracy of the model. Statistics of vibration signals in the time domain or the frequency domain can reveal distinct features of faults in rotating machinery. For example, Xia {\it et al.} used 10 time domain statistical features in Ref.[34] including Absolute mean (Tim-Abm), Variance (Tim-Var), Crest(Tim-Cre), Clearance factor(Tim-Clf), Kurtosis(Tim-Kur), Crest factor(Tim-Crf), Root mean square(Tim-Rms), Pulse factor(Tim-Puf), Skewness(Tim-Ske), Shape factor(Tim-Shf), and 5 frequency domain statistical features including Average frequency(Fre-Afr), Crest(Fre-Cre), Kurtosis(Fre-Kur), Mean energy(Fre-Mea), Variance(Fre-Var). Specifically, we list them in Table 1.\ These features form the feature pool for the shallow model. Generally, shallow models refer to the classical models prior to the deep neural network including SVM, ELM, random forest, K-nearest neighbor (KNN), {\it etc.} In this paper, we choose SVM as a representative to develop our DFD method. It is noteworthy that other shallow models can also be used and the proposed DFD method can be generalized accordingly by selecting candidate models from a pre-defined shallow model pool. We leave it as future work.

{\bf 2.\ Short-time Fourier transform}

As shown in Refs.[30,37],Recent researchers show that spectrogram features of speech are superior to Mel frequency cepstral coefficients (MFCC) with DNN. The spectrogram of the raw vibration wave signal is calculated by applying a short-time Fourier transform (STFT). Mathematically, it can be described as follows:
\begin{small}
\begin{equation}
{{s}_{i}}\left( k,m \right)=\sum\limits_{n=0}^{N-1}{{{x}_{i}}}\left( n \right)w\left( m-n \right){{e}^{-j\frac{2\pi }{N}kn}},
\end{equation}
\end{small}
where ${x_i}$  is the $i^{th}$ channel wave signal of a sample,  $w( \cdot )$ is the window function, {\it e.g.}, Hamming window. ${s_i}\left( {k,m} \right)$ is the spectrogram of ${x_i}$, which has a two-dimension structure.  Fig.4(b) shows the spectrograms of first channel in case 1.  Fig.9 shows the first channel's spectrograms of 9 fault types in the case 2. In this paper, we extract discriminative time-frequency domain features from those spectrograms and use them to train SVM models on the scarce labeled samples.
{\tabcolsep=2.5pt \footnotesize
\begin{center}
  \begin{tabular}{|c|c|c|}
   \multicolumn{3}{c}{\bf Table 1.\ Features selected in different domains}\\ \hline
    Domain & Feature Name & Expression \\  \hline
    \multicolumn{1}{|c|}{\multirow{10}[0]{*}{Time}} & Absolute mean & $\frac{1}{n}\sum\nolimits_{i=1}^{n}{\left| {{x}_{i}} \right|}$ \\
    \multicolumn{1}{|c|}{} & Variance & $\frac{1}{n}\sum\nolimits_{i=1}^{n}{({{x}_{i}}}-x{{)}^{2}}$ \\
    \multicolumn{1}{|c|}{} & Crest & $\max \left( \left| {{x}_{i}} \right| \right)$ \\
    \multicolumn{1}{|c|}{} & Clearance factor & $\max \left( \left| {{x}_{i}} \right| \right)/{{\left( \frac{1}{n}\sum\nolimits_{i=1}^{n}{\sqrt{x_{i}^{2}}} \right)}^{2}}$ \\
    \multicolumn{1}{|c|}{} & Kurtosis & $\frac{1}{n}\sum\nolimits_{i=1}^{n}{x_{i}^{4}}$ \\
    \multicolumn{1}{|c|}{} & Crest factor & $\max \left( \left| {{x}_{i}} \right| \right)/\sqrt{\frac{1}{n}}\sum\nolimits_{i=1}^{n}{x_{i}^{2}}$ \\
    \multicolumn{1}{|c|}{} & Root mean square & $\frac{1}{n}{{\sum\nolimits_{i=1}^{n}{\left( {{X}_{i}}-\bar{X} \right)}}^{2}}$ \\
    \multicolumn{1}{|c|}{} & Pulse factor & $\max \left( \left| {{x}_{i}} \right| \right)/\left( \frac{1}{n}\sum\nolimits_{i=1}^{n}{\left| {{x}_{i}} \right|} \right)$ \\
    \multicolumn{1}{|c|}{} & Skewness & $\frac{1}{n}\sum\nolimits_{i=1}^{n}{x_{i}^{3}}$ \\
    \multicolumn{1}{|c|}{} & Shape factor & $\sqrt{\frac{1}{n}\sum\nolimits_{i=1}^{n}{x_{i}^{2}}}/\left( \frac{1}{n}\sum\nolimits_{i=1}^{n}{\left| {{x}_{i}} \right|} \right)$ \\
    \hline
    \multicolumn{1}{|c|}{\multirow{5}[0]{*}{Frequency}} & Average frequency & $\left( \sum\nolimits_{i=1}^{n}{{{\omega }_{i}}{{X}_{i}}} \right)/\sum\nolimits_{i=1}^{n}{{{X}_{i}}}$ \\
    \multicolumn{1}{|c|}{} & Crest & $\max \left( \left| {{X}_{i}} \right| \right)$ \\
    \multicolumn{1}{|c|}{} & Kurtosis & $\frac{1}{n}\sum\nolimits_{i=1}^{n}{X_{i}^{4}}$ \\
    \multicolumn{1}{|c|}{} & Mean energy & $\frac{1}{n}\sum\nolimits_{i=1}^{n}{{{X}_{i}}}$ \\
    \multicolumn{1}{|c|}{} & Variance & $\frac{1}{n}{{\sum\nolimits_{i=1}^{n}{\left( {{X}_{i}}-\bar{X} \right)}}^{2}}$ \\  \hline
\end{tabular}%
\end{center}}
{\bf 3.\ Convolutional neural network for fault diagnosis}

Convolutional neural network is recently used in fault diagnosis due to its end-to-end modeling capacity. A typical convolutional neural network consists of cascaded convolutional blocks, which are constructed by stacking convolutional layer, batch norm layer$^{[38]}$, non-linear activation layer like ReLU$^{[39]}$ and pooling layer sequentially. Then, several fully-connected layers are used to transform the prior features further and learn a classifier. Prior work leverage 1D CNN to learn features from the raw vibration wave signal directly. However, frequency features are very important for fault diagnosis which can only be learned implicitly from the wave signal. Different from their methods, we feed spectrograms into the specifically designed 2D CNN to learn more discriminative time-frequency domain features in this paper.

\begin{center}{\large\bf III.\ The Proposed Method
}\end{center}

In this paper, we proposed a novel deep CNN method based on knowledge-transferring from shallow models for rotating machinery fault diagnosis with scarce labeled samples. The pipeline of the proposed method is illustrated in Fig.1(a). First, the raw vibration signal is converted into different domains including the original time domain, frequency domain by applying the Fast Fourier transform (FFT) and the time-frequency domain by applying the short-time Fourier transform (STFT). Next, different kinds of features listed in in Table 1\ are extracted accordingly and form a feature candidate pool. Then, we choose SVM as the representative shallow models to train a number of models by selecting different features and their combinations on those given scarce labeled samples. Next, best-performed models are selected to form the pretrained model candidate pool. Then, the proximate labels of those unlabeled samples can be predicted by selecting the appropriate models in the pretrained model candidate pool. Those proximate labels can be regarded as the data format of expert knowledge and are combined together with the scarce fine labeled samples to form the final augmented training set (ATS). Finally, they are used to supervise training a 2-d deep CNN model for better discriminative ability. We'll present each part in detail in the following parts.
\end{multicols}
\vskip 4mm
\centerline{\includegraphics[width=0.8\linewidth]{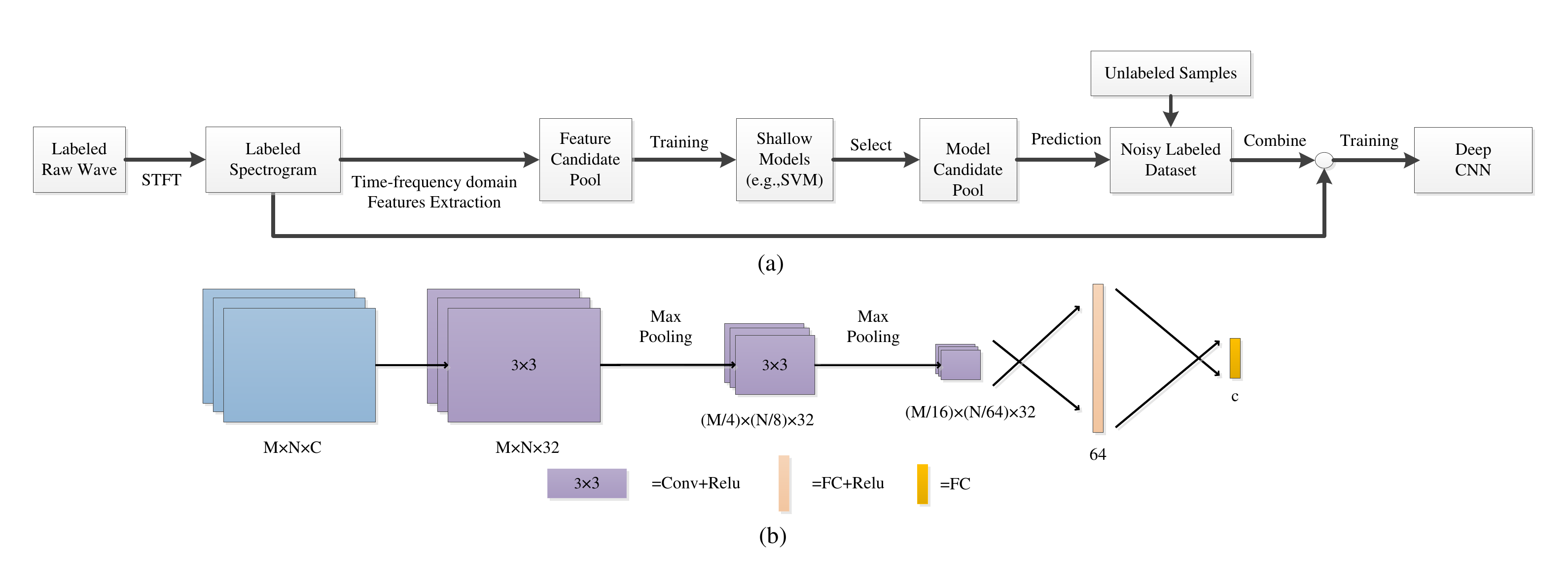} }
\vskip 1mm
\centerline{\footnotesize\begin{tabular}{c} Fig.\ 1.\
(a) Pipeline of the proposed method   (b) Illustration of the proposed deep CNN structure$^{*}$
\end{tabular}}
\vskip 1\baselineskip
\begin{multicols}{2}
\footnotetext{\footnotesize $^{*}$ All the figures in the paper are best viewed in color.}
{\bf 1.\ Constructing feature candidate pool}

First, we follow Ref.[34] to extract the time domain and frequency domain features listed in Table 1.\  Then, considering that the representation in the time-frequency domain can reveal both time and frequency characteristics of fault signals, the raw vibration wave signal is converted into a spectrogram with a two-dimensional structure by using a short-time Fourier transform. The mean, the variance, and the root mean square, {\it etc.}, are calculated along the time axis or the frequency domain axis of the spectrogram to represent the statistical features, which reveal a lot of inherent characteristics of the fault samples. Mathematically, those features can be denoted as follows:
\begin{equation}
{{F}^{t}}=\left\{ f_{i}^{t}\left| i\in {{\Lambda }^{t}} \right. \right\},
\end{equation}
\begin{equation}
{{F}^{fre}}=\left\{ f_{i}^{fre}\left| i\in {{\Lambda }^{fre}} \right. \right\},
\end{equation}
\begin{equation}
{{F}^{tf}}=\left\{ f_{i}^{tf}\left| i\in {{\Lambda }^{tf}} \right. \right\},
\end{equation}
\begin{equation}
{F={{F}^{t}}\bigcup {{F}^{fre}}\bigcup {{F}^{tf}}},
\end{equation}
where $\Lambda ^t$, $\Lambda ^{fre}$ and $\Lambda ^{tf}$ denote the index set of time domain, frequency domain and time-frequency domain features listed in Table 1, respectively. ${f_i^t}$, ${f_i^{fre}}$ and ${f_i^{tf}}$ represent each feature in these domains. ${F_i^t}$, ${F_i^{fre}}$ and ${F_i^{tf}}$ denote the feature set in these domains, and $F$ represents the final feature candidate pool.

{\bf 2.\ Pre-training shallow models on the feature pool}

In this paper, we choose the support vector machine (SVM) as the representative shallow model. Other models like ELM, DT, RF and the complementary between them can also be exploited. We leave it as the future work. We split the scarce labeled samples into training set $D_{T}^{s}$ and validation set ${D_V^s}$ (${{D}^{s}}=D_{T}^{s}\bigcup D_{V}^{s}$). Given the feature candidate pool $F$ and the scarce labeled samples in ${D_T^s}$ , several SVM models can be trained. Mathematically, it can be formulated as follows:

\begin{equation}
M=\left\{ {{m}_{i}}\left( {{F}_{i}} \right)\left| {{F}_{i}}\subseteq F \right. \right\},
\end{equation}
where ${{m}_{i}}\left( \cdot  \right)$ denotes the ${{i}^{th}}$ trained model based on a feature combination ${{F}_{i}}$:

\begin{equation}
{{F}_{i}}\text{=}{\mathop{\bigcup }}_{j\in \Lambda _{i}^{F}}\,{{f}_{ij}},
\end{equation}
where $\Lambda _{i}^{F}$ denotes the ${{i}^{th}}$ index set of ${{F}_{i}}$. Then, we choose the top-k performed models from $M$ to form the model candidate pool ${{M}^{C}}$
\begin{equation}
{{M}^{C}}=\left\{ {{m}_{i}}\left| Rank\left( Acc\left( {{m}_{i}} \right) \right)\in \left\{ 0,1,\ldots ,k-1 \right\} \right. \right\},
\end{equation}
where $Acc\left( {{m}_{i}} \right)$ denotes the fault classification accuracy of model ${{m}_{i}}$ on the validation set $D_{V}^{s}$. Next, we can make predictions on the unlabeled samples by leveraging the selected pre-trained SVM model from ${{M}^{C}}$:
\begin{equation}
{{D}^{u}}=\left\{ \left( {{x}_{i}},{{y}_{i}} \right)\left| i\in {{\Lambda }^{u}} \right. \right\},
\end{equation}
where ${{y}_{i}}$ is the prediction of sample ${{x}_{i}}$ in the unlabeled sample set, ${{\Lambda }^{u}}$ is the index set of ${{D}^{u}}$. ${{y}_{i}}$ is be calculate by fusing predictions from different candidate models:
\begin{equation}
{{y}_{i}}=g\left( {{m}_{k}}\left( {{x}_{i}} \right)\left| {{m}_{k}}\in {{M}^{C}} \right. \right),
\end{equation}
where $g\left( \cdot  \right)$  represents the fusion function. In this paper, we calculate the arithmetical mean of the predicted probabilities from different models and selected the category with the highest probability as the final fusion result. ${{D}^{u}}$ is called the quasi-labeled dataset in this paper. Fig.2 illustrates the above process.
\end{multicols}
\vskip 4mm
\centerline{\includegraphics[width=0.55\linewidth]{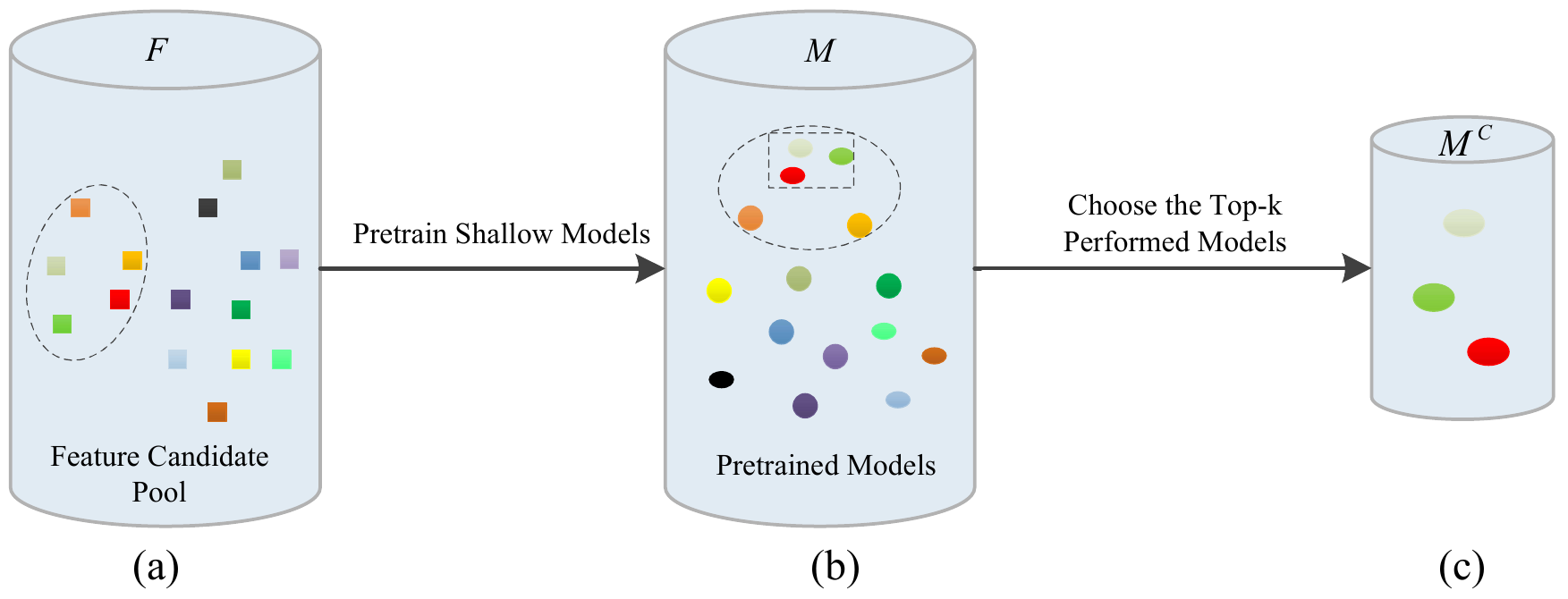}}
\vskip 1mm
\centerline{\footnotesize\begin{tabular}{c} Fig.\ 2.\
(a)The feature candidate pool $F$  (b) The pretrained model on $F$  (c) The model candidate pool
\end{tabular}}
\begin{multicols}{2}
{\bf 3.\ The proposed deep CNN architecture}

Different from prior work which uses the raw vibration wave signal as input, we leverage its spectrograms as input which has good representation ability in both time and frequency domain. Therefore, we design a 2D CNN based deep network structure. Specifically, the proposed architecture consists of 2 convolutional blocks and 2 fully-connected layers. Each convolutional block consists of convolutional layers, nonlinear activation layer, {\it i.e.}, ReLU in this paper, and is followed by a max pooling layer. The convolution process is designed to capture time-frequency features from the input spectrograms. We use the cross-entropy loss and mini-batch stochastic gradient descent algorithm to train the network. The illustration of the proposed deep CNN architecture is shown in Fig.1(b). The details of the deep CNN architecture designed according to the dataset in case 1 are shown in in Table 2\  and it contains 20,160 parameters and has 1.06x10$^7$ flops computation. Again, the network is also used in case 2.
{\tabcolsep=2.0pt \footnotesize
\begin{center}
\begin{tabular}{|p{10.75em}<{\centering}|c|c|c|c|c|}
\multicolumn{6}{c}{\bf Table 2.\ Network architectures of the proposed CNN}\\ \hline
    Type  & Input Size & Number   & Filter & Pad   & stride \\ \hline
    Conv1 & 32x128x8 & 32   & 3x3   & 1     & 1 \\     \hline
    Pool1 & 32x128x32 & -     & 4x8     & 0     & (4,8) \\     \hline
    Conv2 & 8x16x32 & 32     & 3x3   & 1     & 1 \\     \hline
    Pool2 & 8x16x32 & -    & 4x8   & 0     & (4,8) \\      \hline
    Fc3 & 2x2x32 & 64     & -   & -     & - \\     \hline
    Fc4 & 64x1x1 & 7     & -   & -     & - \\     \hline
    Params & \multicolumn{5}{c|}{20,160}         \\     \hline
    Complexity$^{*}$  & \multicolumn{5}{c|}{1.06x10$^7$}         \\  \hline
\end{tabular}%
\end{center}}
\footnotetext{\footnotesize $^{*}$The complexity of the proposed deep CNN is calculated only for convolutional layers.}
Given the network, we can directly train it on the scarce labeled samples in $D^{s}$. However, it is prone to overfitting due to the limited training samples. To tackle this issue, we proposed a knowledge-transferring based training method which leverages both the quasi-labeled dataset $D^{u}$ and the original scarce labeled dataset $D^{s}$. We'll present the details in the next part.

{\bf 4.\ Knowledge-transferring from the shallow models to the deep CNN}

As we know that hand-crafted features are designed according to the prior knowledge and diagnostic expertise. They reveal some intrinsic characteristics of fault samples. Different kinds of features reflect specific domain knowledge, {\it i.e.}, time-domain, frequency domain and time-frequency domain. Therefore, the predicted labels in $D^{u}$ can be regarded as the data format of expert knowledge which has been modeled by the SVM models trained on different combinations of those features. Inspired by this observation, we leverage the quasi-labeled dataset set $D^{u}$ to take advantages of the transferred knowledge from the shallow models and hand-crafted features. Consequently, we combine $D^{u}$ and ${D_T^s}$ together to form the final augmented training dataset to train the proposed deep CNN in Section III.3. We suppose that the network can benefit from the augmented samples and learn more discriminative features and a stronger classifier, to achieve the goal that transferring knowledge from the shallow models to the deep CNN can lead to a better model. Mathematically, the training processing can be formulated as minimizing the following cross-entropy loss:
\begin{equation}
{{\theta }^{\text{*}}}\text{=}{\mathop{\arg \min }}_{\theta} \,-    \sum\limits_{i\in {{D}^{u}}
\bigcup D_{T}^{s}}{\sum\limits_{c=0}^{C-1}{{{q}_{ic}}}}\log ({{p}_{ic}}),
\end{equation}
where $\theta$ represents the learnable parameters in the proposed network, $C$ is the number of categories,${{q}_{i}}\in {{R}^{C}}$ is the one-hot vector representation of ground truth label ${{y}_{i}}$, {\it i.e.}, ${{q}_{ic}}=\delta ({{y}_{i}},c)\in \{0,1\}$,${{p}_{i}}\in {{R}^{C}}$ is the predicted probability vector by the network.

\

\begin{center}{\large\bf IV.\ Experiments
}\end{center}

To evaluate the effectiveness of the proposed approach, we conducted extensive experiments on two rotating machinery fault datasets. On each dataset, the experiments include the following three parts: feature selection, pre-training SVM models and evaluating the knowledge-transferring based deep CNN model, which corresponds to the three sections in Section III. We compared the proposed knowledge-transferring based DFD model with the state-of-the-art models as well as the vanilla one which was trained with only the scarce samples. Finally, we analyze the computational complexity.

{\bf1.\ Case one: rotating machinery fault diagnosis}

1)\ Dataset preparation and parameter settings

In this case, we used the ZHS-2 Multi-functional motor platform with flexible rotors as shown in Fig.3 to construct the rotating machinery fault diagnosis dataset. Eight vibrating sensors were installed at the pedestal to record vibration signals. These signals were collected by the HG8902 data collection box. In this experiment, we considered six types of faults: Rotor Unbalance I (RU1), Rotor Unbalance III (RU3), Rotor Unbalance V (RU5), Rotor Unbalance VII (RU7), Pan Page Break (PPB) and Pedestal Looseness (PL) and the Normal condition (N). The first four types of faults were simulated by installing different numbers of screws on the rotor as shown in Fig.3-C, {\it e.g.}, five screws for Rotor Unbalance V. The Pan Page Break was simulated by installing a pan with a broken page on the roller as shown in Fig.3-B. The Pedestal Looseness was simulated by loosening the bolts of the pedestal as shown in Fig.3-A.
\vskip 4mm
\centerline{\includegraphics[width=1\linewidth]{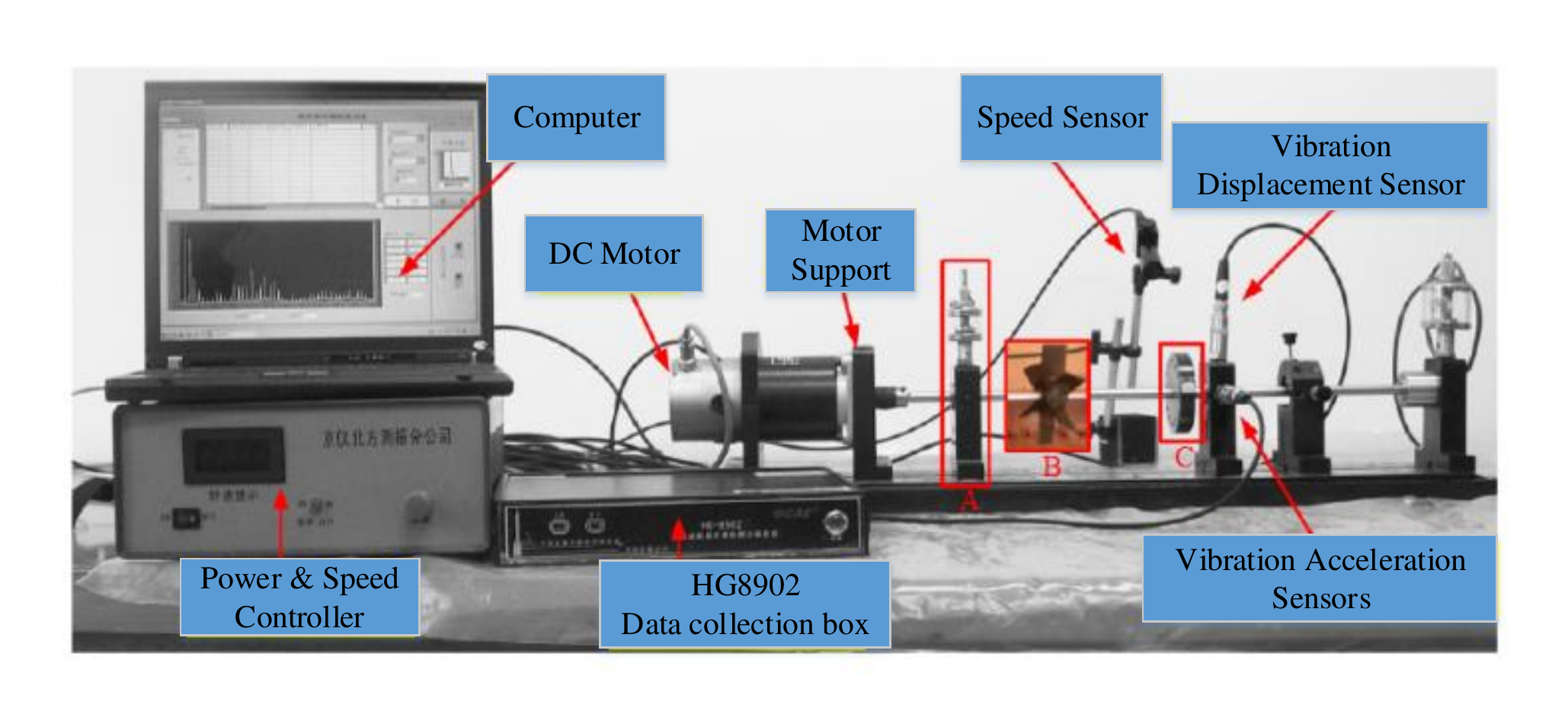} }
\vskip 1mm
\centerline{\footnotesize\begin{tabular}{c} Fig.\ 3.\
Experimental setup in case 1. A: Pedestal, B: Pan, C: Rotor
\end{tabular}}
\vskip 0.5\baselineskip
The rotation speed of the motor rotor was set to 1500r/min. The sampling frequency was 1280Hz. Each sample lasted 8s and 10240 data points were recorded. We collected 300 samples for each fault type and the normal condition. Therefore, 2100 samples were collected in total to form the rotating machinery fault diagnosis dataset. Exemplar waveforms from the first sensor for the aforementioned six types of rotating machinery faults are shown in Fig.4(a).

Moreover, to illustrate the influence of the number of labeled samples, we evenly sample different proportions of samples for each category from the original dataset to form the new datasets. Four datasets with different volumes are constructed, {\it i.e.}, 2\%, 4\%,6\%,8\%.  Then, we train SVMs for each of the datasets and using different features. For each configuration, we use 2 out of 3 splits as the training set and the left one split as the test set.

Parameters are set as the following: We use a Hamming window of length 256 in the short-time Fourier transform with an overlap size of 128, which leads to 2D images of 32*128. Fig.4(b) shows the spectrograms corresponding to the above waveform. The waveform signals collected from the eight sensors are processed as input to the convolutional neural network and the input size is 32*128*8. The CNN model is trained in a total of 10,000 iterations with a batch size of 256. The update strategy of learning rate is set to ``step'', which changes the learning rate every 2,500 iterations and the learning rate decreases by half from 0.01 to 6.25x10$^{-4}$. The momentum and the decay parameter are set to 0.9 and 5x10$^{-6}$, respectively. We implemented the proposed method in CAFFE $^{[40]}$. All the experiments are conducted on a workstation with Nvidia GTX Titan X GPUs if not specified. Besides, the performance metric is average classification accuracy, which is defined as follows:
\begin{equation}
Accuacy=\sum\limits_{i=1}^{i=C}{{{{c}_{ii}}}/{\sum\limits_{i=1}^{i=C}{\sum\limits_{J=1}^{j=C}{{{c}_{ij}}}}}\;}
\end{equation}
Here,${c_{ij}}$ is the number of samples which belong to the $i^{th}$ category and are predicted to the $j^{th}$ category. $C$ is the number of categories.
\vskip 4mm
\centerline{\includegraphics[width=1\linewidth]{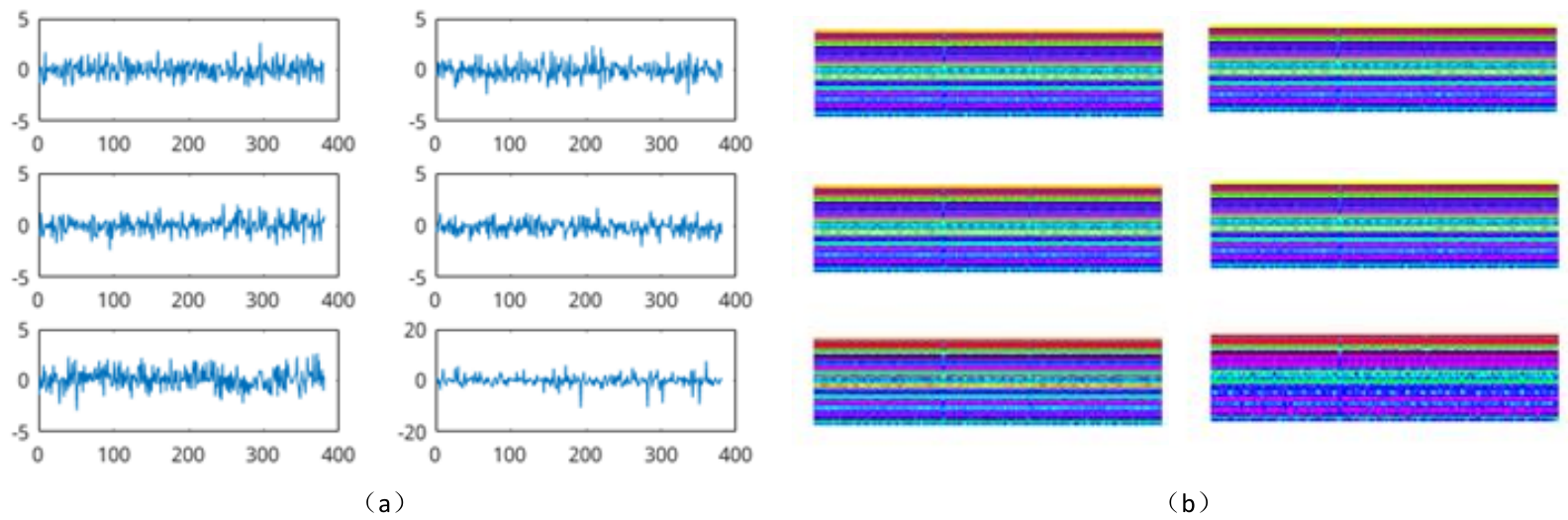}}
\vskip 1mm
\centerline{\footnotesize\begin{tabular}{c} Fig.\ 4.\
(a) Exemplar waveforms  (b) Exemplar spectrograms
\end{tabular}}
\vskip 0.5\baselineskip
2)\ Feature selection for SVM

We trained the SVM model with different volumes of dataset and features. Table 3\ shows the results. As can be seen, the features extracted along the frequency axis are more discriminative than the features extracted along the time axis for all the training sets and test sets. Besides, with the increase of the dataset volume, the performance of Fre-Cre is consistently boosted. The accuracy of Fre-Mea is significantly higher than other features, and the result is true no matter which data set.
{\tabcolsep=2.0pt \footnotesize
\begin{center}
\begin{tabular}{|p{5.5em}<{\centering}|c|c|c|c|c|c|c|c|}
\multicolumn{9}{c}{\bf Table 3.\ Classification accuracy of SVM in case 1}\\ \hline
    \multicolumn{1}{|p{5.5em}<{\centering}|}{\multirow{2}[4]{*}{Feature}} & \multicolumn{2}{c|}{2\%} & \multicolumn{2}{c|}{4\%} & \multicolumn{2}{c|}{6\%} & \multicolumn{2}{c|}{8\%} \\
\cline{2-9}   \multicolumn{1}{|p{5.5em}|}{} & train & test & train & test & train & test & train  & test \\  \hline
    Tim-Abm & 0.804 & 0.571 & 0.789 & 0.679 & 0.754 & 0.575 & 0.772 & 0.595 \\  \hline
    Tim-Var & 0.726 & 0.524 & 0.673 & 0.637 & 0.623 & 0.615 & 0.667 & 0.586 \\  \hline
    Tim-Cre & 0.738 & 0.643 & 0.810 & 0.720 & 0.835 & 0.659 & 0.907 & 0.696 \\  \hline
    Tim-Clf & 0.143 & 0.238 & 0.286 & 0.274 & 0.381 & 0.262 & 0.381 & 0.274 \\  \hline
    Tim-Kur & 0.738 & 0.560 & 0.720 & 0.685 & 0.587 & 0.536 & 0.601 & 0.637 \\  \hline
    Tim-Crf & 0.702 & 0.238 & 0.845 & 0.500 & 0.901 & 0.516 & 0.748 & 0.640 \\  \hline
    Tim-Rms & 0.804 & 0.583 & 0.788 & 0.708 & 0.712 & 0.591 & 0.770 & 0.613 \\  \hline
    Tim-Puf & 0.792 & 0.500 & 0.845 & 0.619 & 0.829 & 0.595 & 0.706 & 0.693 \\  \hline
    Tim-Ske & 0.655 & 0.429 & 0.765 & 0.714 & 0.643 & 0.627 & 0.631 & 0.521 \\  \hline
    Tim-Shf & 0.643 & 0.560 & 0.827 & 0.500 & 0.796 & 0.560 & 0.732 & 0.512 \\  \hline
    Fre-Afr & 0.000 & 0.143 & 0.000 & 0.143 & 0.000 & 0.143 & 0.000 & 0.143 \\  \hline
    Fre-Cre & 0.988 & 0.833 & 0.988 & 0.940 & 0.988 & 0.944 & 1.000 & 0.988 \\  \hline
    Fre-Kur & 0.929 & 0.762 & 0.935 & 0.929 & 0.968 & 0.897 & 0.982 & 0.943 \\  \hline
    Fre-Mea & 1.000 & 1.000 & 0.988 & 0.988 & 1.000 & 1.000 & 1.000 & 0.994 \\  \hline
    Fre-Var & 0.994 & 0.976 & 0.994 & 0.988 & 1.000 & 1.000 & 1.000 & 0.991 \\  \hline
      Fuse  & 1.000 & 1.000 & 1.000 & 1.000 & 1.000 & 1.000 & 1.000 & 1.000 \\  \hline
\end{tabular}%
\end{center}}
To show it more clearly, we use visualization techniques$^{[41]}$ to shed light on the differences between features. First, we calculate the features Tim-Abm and Fre-Mea of the samples in the dataset and made predictions using the trained SVM models. Then, the visualization of the features by using t-SNE $^{[41]}$ is shown in Fig.5. As can be seen from Fig.5(a), predictions from Rotor Unbalance I are confused with Normal condition and Pan Page Break. It implies that Rotor Unbalance I and Pan Page Break are two minor faults and it is difficult to distinguish among Rotor Unbalance I, Pan Page Break and the normal condition. Therefore, more discriminative features should be utilized to distinguish the above three conditions. As can be seen from Fig.5(b), Fre-Mea can tackle this challenge and achieve better performance which is consistent with the results in Table 3.\ Besides, predictions from Rotor Unbalance III are confused with Rotor Unbalance V. Again, Fre-Mea achieves better results than Tim-Abm as shown in Fig.5(b).
\vskip 4mm
\centerline{\includegraphics[width=1\linewidth]{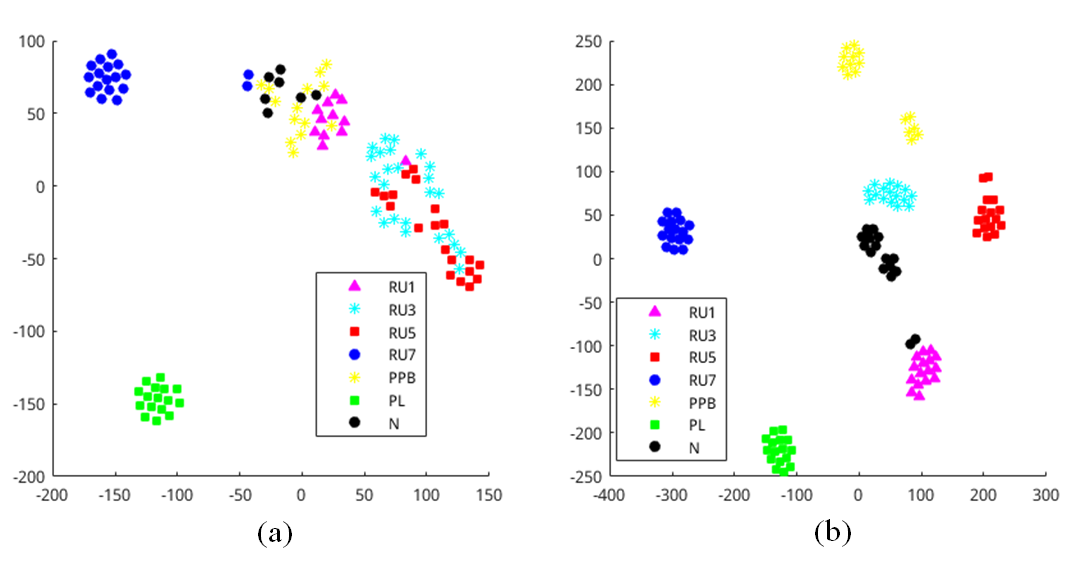}}
\vskip 1mm
\centerline{\footnotesize\begin{tabular}{c} Fig.\ 5.\
(a) Visualization of Tim-Var  (b) Visualization of Fre-Var
\end{tabular}}
\vskip 0.5\baselineskip
Each of the pre-trained 15 models using different features reflects different aspects of the expert knowledge and experience, which can be combined to take advantage of their complementary and achieve better results. According to the classification accuracy, we choose the SVM model trained using Fre-Var and Fre-Mea to form the candidate model pool. We adopt simple arithmetic mean of the predicted probability vectors from both models as the final fusion prediction. The training and test accuracy are shown in the last row of Table 3.\ In all datasets, the classification accuracy is boosted to 100\%.

3)\ Knowledge-transferring for deep CNN

We trained the proposed deep CNN model with different dataset to show the influences of data volumes. The results are shown in  Table 4.\ With the increase of dataset volume, its performance becomes better and better. It shows that the model capacity of deep CNN can be exploited with more training samples.

Then, the deep CNN model was retrained on the ATS to take advantage of the transferred knowledge from the shallow model in the model candidate pool. The results are shown in the bottom rows of  Table 4.\ It can be seen that with the help of augmented dataset, the proposed deep CNN learns more discriminative features and achieve better classification accuracy than both the vanilla deep CNN and the corresponding pre-trained SVM models. It implies that knowledge from shallow models can be transferred to the deep CNN model in a manner of predicted sample labels.
{\tabcolsep=2.5pt \footnotesize
\begin{center}
\begin{tabular}{|p{15.19em}<{\centering}|c|c|c|c|}
\multicolumn{5}{c}{\bf Table 4.\ Accuracy of the models}\\ \hline
    Dataset Volumn & 2\%   & 4\%   & 6\%   & 8\% \\    \hline
    CNN(Train)& 1     & 1     & 1     & 1 \\   \hline
    CNN(Test) & 0.928 & 0.964 & 0.976 & 0.982 \\   \hline
    CNN-T(Fre-Mea;Train) & 1     & 1     & 1     & 1 \\   \hline
    CNN-T(Fre-Mea;Test) & 1     & 1     & 1     & 1 \\   \hline
    CNN-T-Fsuion(Train) & 1     & 1     & 1     & 1 \\  \hline
    CNN-T-Fusion(Test) & 1     & 1     & 1     & 1 \\  \hline
\end{tabular}%
\end{center}}

4)\ Running time analysis

The inference times of the proposed deep CNN at different modes (GPU and CPU) with different settings of batch size are plotted in Fig.6.  As can be seen, the average inference time for each sample almost keeps the same with the increase of batch size at the CPU mode. As for the GPU mode, it is decreased consistently. A 20x speedup ratio is achieved by leveraging the parallel acceleration of modern GPUs. For example, a batch of 1024 samples can be processed in less than 0.1 seconds on the NVIDIA GTX 1080ti GPU, while costing more than 2 seconds on the Intel Core i7 CPU.
\vskip 4mm
\centerline{\includegraphics[width=0.7\linewidth]{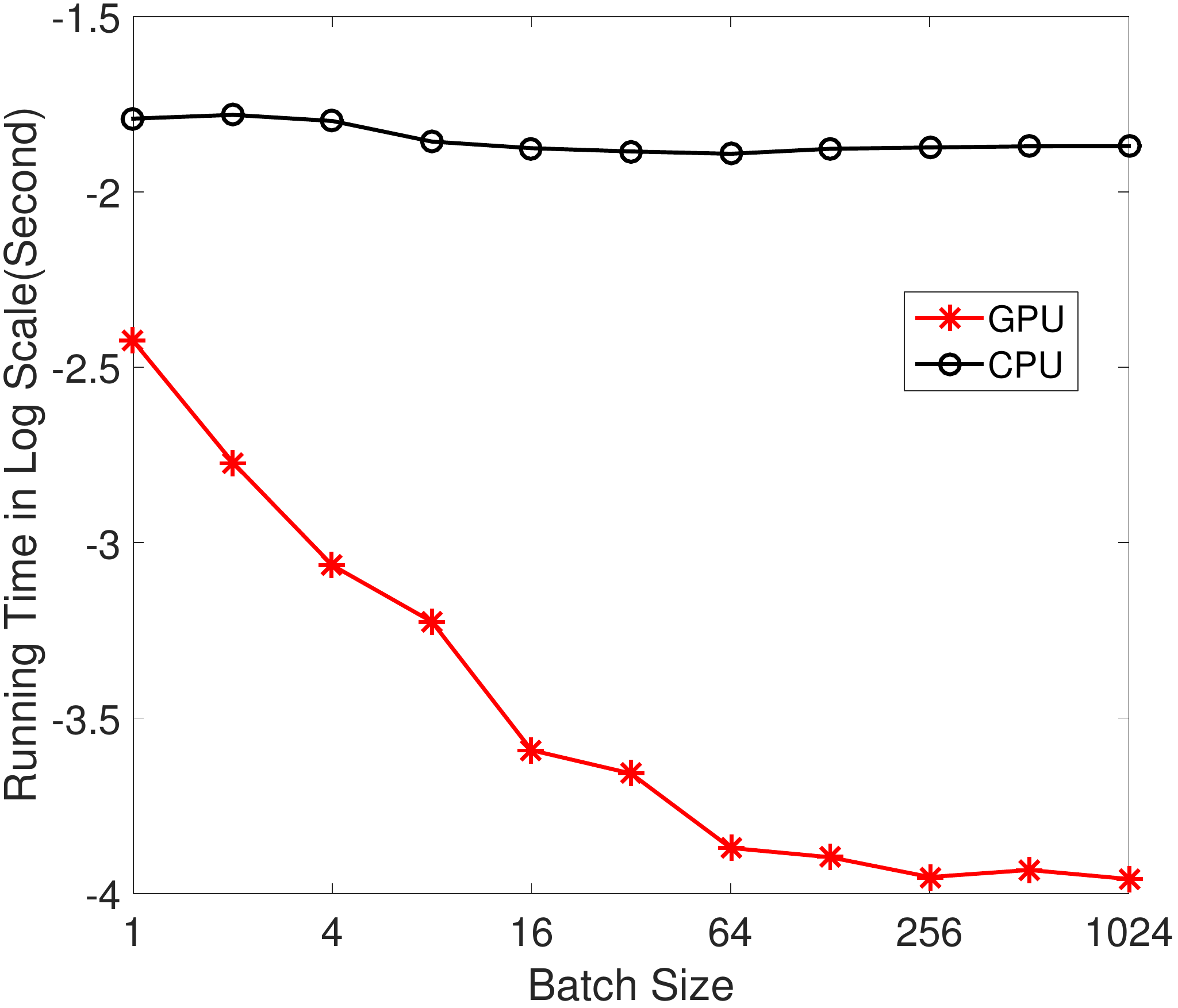}}
\vskip 1mm
\centerline{\footnotesize\begin{tabular}{c} Fig.\ 6.\
The inference times with different settings of batch size in case 1
\end{tabular}}
\vskip 0.5\baselineskip

{\bf2.\ Case two: bearing fault diagnosis}

1)\ Dataset preparation and parameter settings

In this case, the publicly available roller bearing condition dataset from Case Western Reserve University (CRWU) is analyzed$^{[42,43]}$. Fig.7\ shows the test platform. Vibration signals are collected using sensors mounted in three different locations: the drive end, the fan end, and the base. In the experiment, we used data with 1 horsepower motor load, considering 9 types of fault: ball defect I (B007), ball defect II (B014), ball failure defect(B021), inner race defect I (IR007), inner race defect II (IR014), inner race defect III (IR021) and outer race defect I (OR007), outer race defect OR014), outer race defect III (OR021). A single point fault was introduced to each bearing by electrodischarge machining with fault diameters of 0.07, 0.14, and 0.21miles$^{[34]}$.
\vskip 4mm
\centerline{\includegraphics[width=0.93\linewidth]{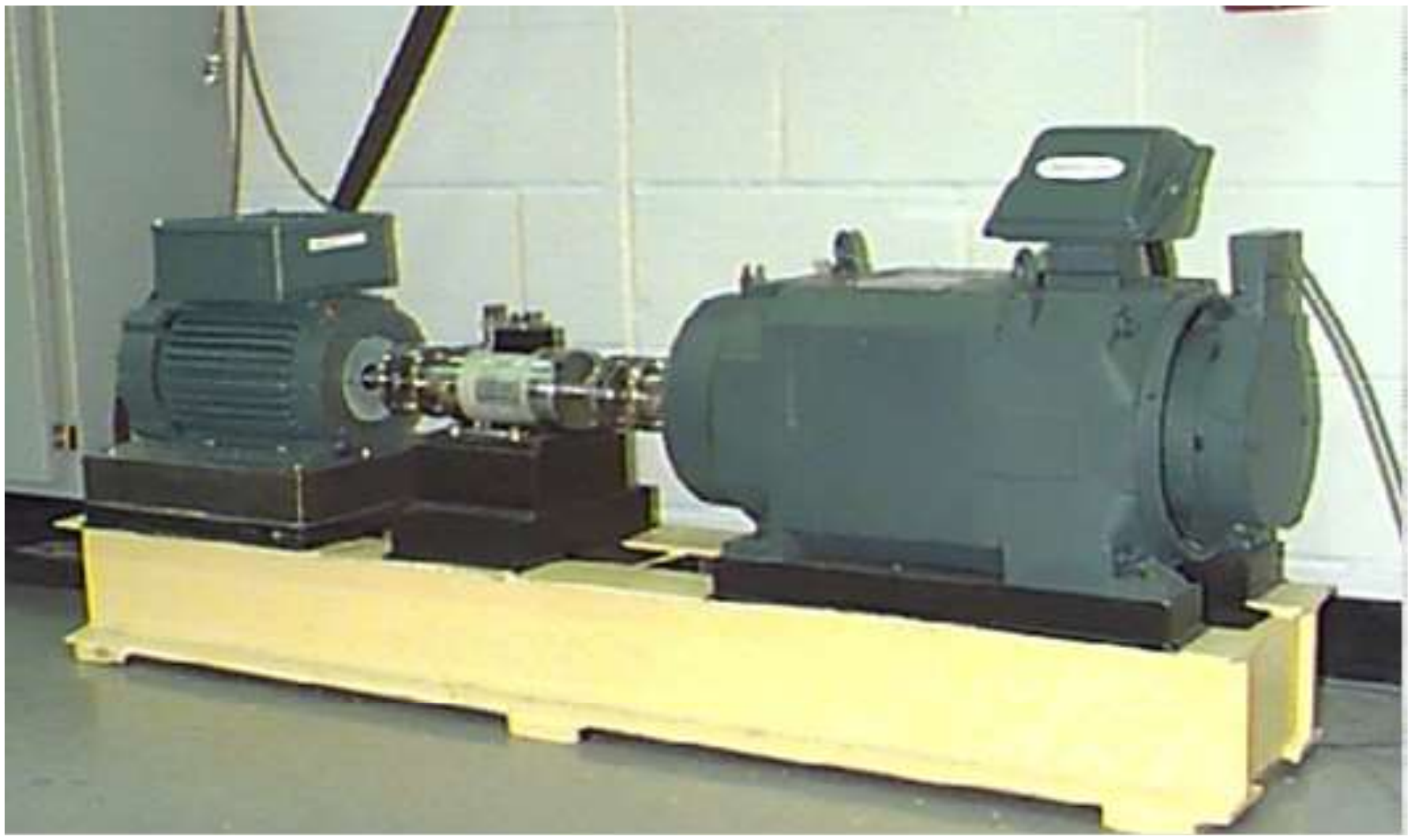}}
\vskip 1mm
\centerline{\footnotesize\begin{tabular}{c} Fig.\ 7.\
The test platform of Bearing Fault dataset
\end{tabular}}
\vskip 0.5\baselineskip

The rotation speed of the motor rotor was set to 1772r/min. The sampling frequency was 12KHz. Approximately 400 signals can be acquired per revolution of the bearing, so continuous 400 signals can be defined as one sample. Therefore, 2700 samples were collected in total to form the dataset. Exemplar waveforms from the first sensor for the aforementioned nine types of faults are shown in Fig.8.\
\vskip 4mm
\centerline{\includegraphics[width=1\linewidth]{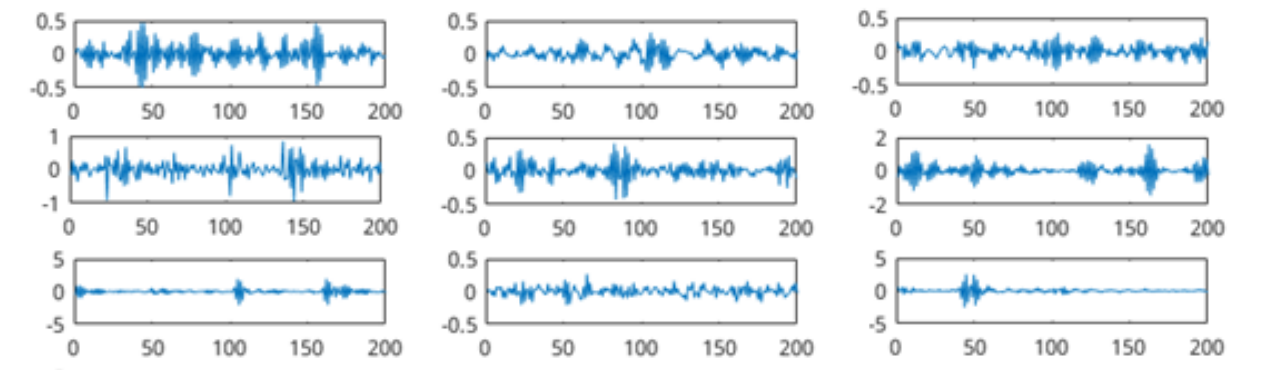}}
\vskip 1mm
\centerline{\footnotesize\begin{tabular}{c} Fig.\ 8.\
Exemplar waveforms from the first sensor
\end{tabular}}
\vskip 0.5\baselineskip
Parameters are set as the following. We use a Hamming window of length 16 in the short-time Fourier transform with an overlap size of 8, due to the small number of sample points, which lead to a 8*64 2-D image. Fig.9 shows the spectrograms corresponding to the above waveform. The waveform signals collected from the three sensors at the driving end, the fan end, and the pedestal are processed as input to the convolutional neural network and the input size is 8*64*3. The rest of the parameter is the same as parameters set in Case 1.
\vskip 4mm
\centerline{\includegraphics[width=0.93\linewidth]{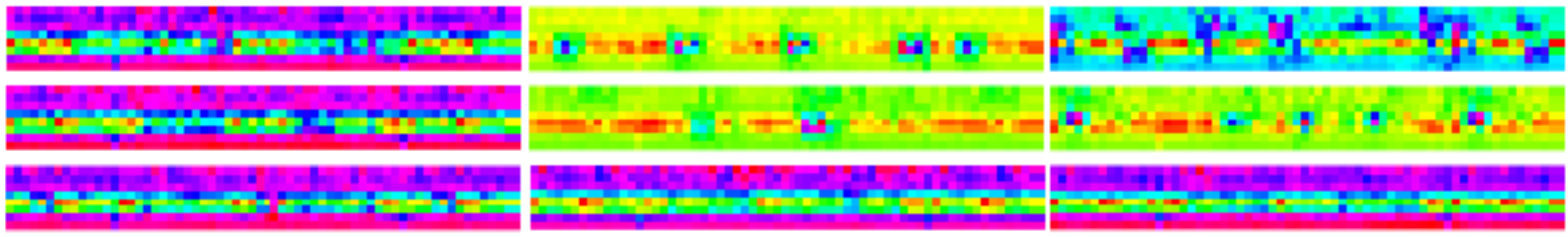}}
\vskip 1mm
\centerline{\footnotesize\begin{tabular}{c} Fig.\ 9.\
The spectrograms corresponding to the above waveform
\end{tabular}}
\vskip 0.5\baselineskip
2)\ Feature selection for SVM

The features are calculated respectively along the time axis and the frequency domain axis and the SVM models are trained with different sample volumes and features. The results are shown in the Table 5.\ As can be seen, there is a serious over-fitting phenomenon with the features extracted along time axis. The accuracy of the training dataset is much higher than the accuracy of the test dataset. As for the features extracted along frequency axis, they are much more discriminative than features extracted along time axis. Again, the most discriminative feature is still Fre-Mea.

The differences between the features are further elaborated by using t-SNE. First, we calculate the features Tim-var and Fre-Mea of the samples in the dataset and made predictions using the trained SVM models. As can be seen from Fig.10(a), B007 is relatively easy to distinguish, and the predictions of B014 and B021 are confused with OR014. The predictions of the remaining five categories are confusing, indicating that feature Tim-var is difficult to distinguish the three levels of inner ring defect. Fortunately, Fre-Mea can tackle this challenge and achieve better performance.
\vskip 4mm
\centerline{\includegraphics[width=0.93\linewidth]{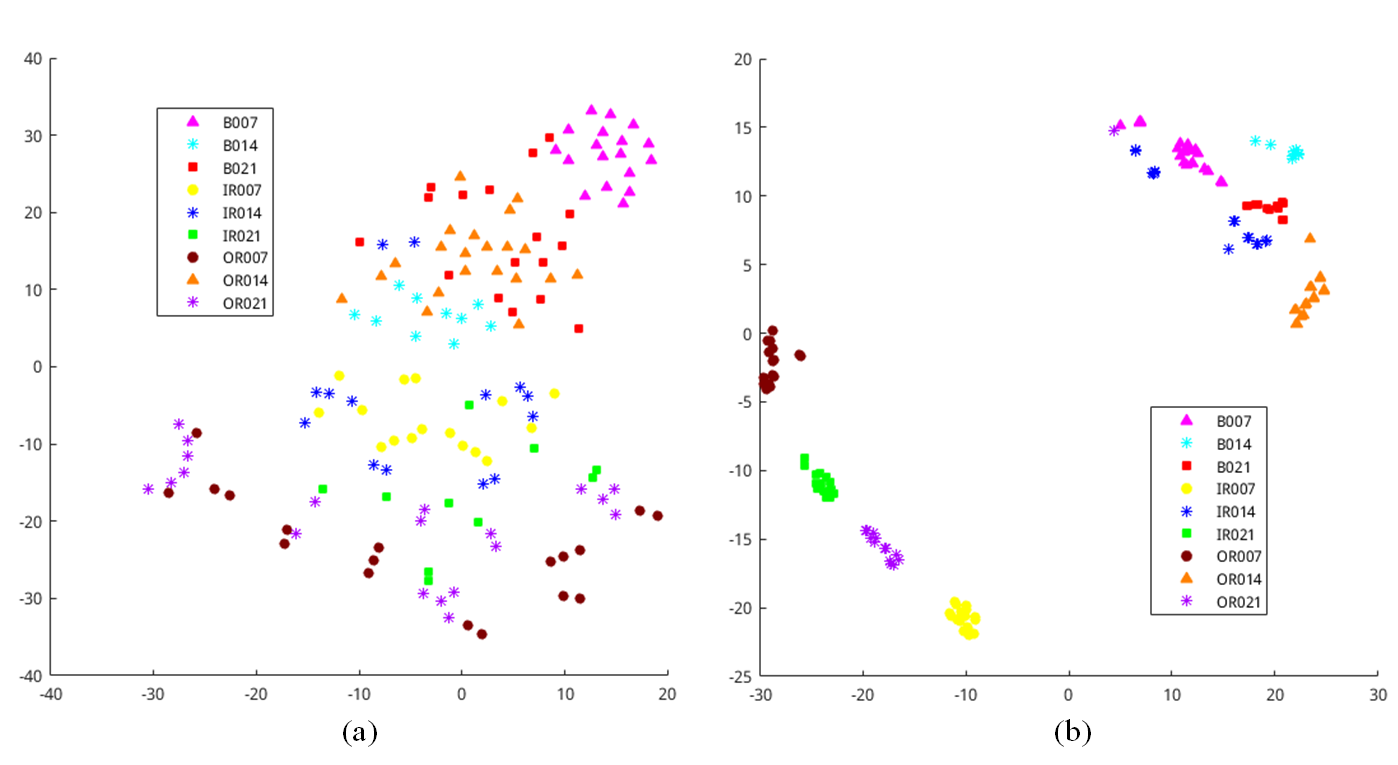}}
\vskip 1mm
\centerline{\footnotesize\begin{tabular}{c} Fig.\ 10.\
a) Visualization of Tim-Var  (b) Visualization of Fre-Mea
\end{tabular}}
According to the classification accuracy, the SVM model trained using Fre-Var and Fre-Mea is selected to form the candidate model pool. We deduce that Fre-Var and Fre-Mea are suitable hand-crafted features for distinguishing fault types from rotating machinery. The prediction labels of the above two models are combined, and the results are shown in the last raw of Table 5.\ Generally, the fusion results are better than each of the single one.
{\tabcolsep=2.0pt \footnotesize
\begin{center}
\begin{tabular}{|p{5.5em}<{\centering}|c|c|c|c|c|c|c|c|}
\multicolumn{9}{c}{\bf Table 5.\ Classification accuracy of SVM in case 2}\\ \hline
    \multicolumn{1}{|p{5.5em}<{\centering}|}{\multirow{2}[4]{*}{Feature}} & \multicolumn{2}{c|}{2\%} & \multicolumn{2}{c|}{4\%} & \multicolumn{2}{c|}{6\%} & \multicolumn{2}{c|}{8\%} \\
\cline{2-9}   \multicolumn{1}{|p{5.5em}|}{} & train & test & train & test & train & test & train  & test \\  \hline
    Tim-Abm & 0.755 & 0.269 & 0.815 & 0.528 & 0.846 & 0.596 & 0.860 & 0.535 \\  \hline
    Tim-Var & 0.787 & 0.500 & 0.951 & 0.671 & 0.846 & 0.583 & 0.878 & 0.576 \\  \hline
    Tim-Cre & 0.745 & 0.481 & 0.914 & 0.579 & 0.866 & 0.608 & 0.905 & 0.588 \\  \hline
    Tim-Clf & 0.000 & 0.111 & 0.111 & 0.111 & 0.111 & 0.111 & 0.111 & 0.111 \\  \hline
    Tim-Kur & 0.495 & 0.398 & 0.560 & 0.421 & 0.472 & 0.414 & 0.497 & 0.414 \\  \hline
    Tin-Crf & 0.648 & 0.352 & 0.653 & 0.324 & 0.665 & 0.247 & 0.774 & 0.269 \\  \hline
    Tim-Rms & 0.819 & 0.426 & 0.880 & 0.542 & 0.847 & 0.605 & 0.873 & 0.558 \\  \hline
    Tim-Puf & 0.685 & 0.231 & 0.752 & 0.361 & 0.610 & 0.333 & 0.922 & 0.384 \\  \hline
    Tim-Ske & 0.352 & 0.389 & 0.491 & 0.417 & 0.505 & 0.426 & 0.543 & 0.451 \\  \hline
    Tim-Shf & 0.588 & 0.296 & 0.755 & 0.333 & 0.616 & 0.336 & 0.707 & 0.326 \\  \hline
    Fre-Afr & 0.000 & 0.111 & 0.000 & 0.111 & 0.000 & 0.111 & 0.000 & 0.111 \\  \hline
    Fre-Cre & 0.917 & 0.722 & 0.894 & 0.750 & 0.873 & 0.716 & 0.866 & 0.755 \\  \hline
    Fre-Kur & 0.731 & 0.398 & 0.806 & 0.532 & 0.810 & 0.565 & 0.829 & 0.549 \\  \hline
    Fre-Mea & 0.991 & 0.935 & 0.970 & 0.944 & 0.991 & 0.969 & 0.981 & 0.954 \\  \hline
    Fre-Var & 0.843 & 0.750 & 0.847 & 0.769 & 0.821 & 0.759 & 0.829 & 0.759 \\  \hline
    Fuse    & 0.991 & 0.963 & 0.970 & 0.954 & 0.991 & 0.953 & 0.981 & 0.956 \\  \hline
\end{tabular}%
\end{center}}

3)\ Knowledge-transferring for deep CNN

The deep CNN architecture for Bearing Fault Dataset is same as the structure of the first case. We trained the proposed vanilla deep CNN model with the scarce labeled samples. The results are shown in Table 6.\ The results of deep CNN model retrained on ATS are shown in the bottom rows of Table 6.\ It can be seen that the accuracy of the CNN model is improved through the knowledge-transferring based method. It is noteworthy that Fre-var is not so discriminative as in the first case. Generally, it leads to inferior accuracy than Fre-Mea, which degrades the proposed DFD model based on the fusion labels. It implies that carefully selecting the best features and pre-trained models based on their validation accuracies is important for the final performance of the proposed DFD. Xia {\it et al.} proposed a deep CNN method using multiple sensors. The result of experiment about the dataset from Case Western Reserve University is 99.44\%$^{[34]}$. However, the proposed method in this paper can reach an accuracy of 98.4\% with scarce labeled samples ({\it i.e.}, only 8\% labeled samples are used). This method is of great significance in tackling the label scarcity issue.
{\tabcolsep=2.5pt \footnotesize
\begin{center}
\scalebox{0.83}{
\begin{tabular}{|p{10.65em}<{\centering}|c|c|c|c|c|c|c|c|}
 \multicolumn{9}{c}{\bf Table 6.\ Classification results of models}\\ \hline
   \multicolumn{1}{|c}{\multirow{2}[2]{*}{Dataset Volumn}} & \multicolumn{2}{|c}{2\%} & \multicolumn{2}{|c}{4\%} & \multicolumn{2}{|c}{6\%} & \multicolumn{2}{|c|}{8\%} \\
    \cline{2-9}    \multicolumn{1}{|c|}{} & Ave. & Std. & Ave. & Std. & Ave. & Std. & Ave. & Std. \\ \hline
    CNN(Train) & 1  &-   & 1  &-    & 1  &-    & 1 &-   \\ \hline
    CNN(Test) & 0.907 & 0.035  & 0.944 & 0.022  & 0.963 & 0.013 & 0.97  &0.012 \\ \hline
    CNN-T(Fre-Mea;Train) & 1  &-   & 1  &-   & 1   &-  & 1  &- \\ \hline
    CNN-T(Fre-Mea;Test) & 0.907 & 0.035 & 0.958 &0.011 & 0.969 & 0.017 & 0.984  &0.013 \\ \hline
    CNN-T-Fusion(Train) & 1 &-    & 1  &-   & 1  &-   & 1 &- \\ \hline
    CNN-T-Fusion(Test) & 0.907 & 0.057 & 0.958 & 0.022 & 0.966 & 0.016 & 0.974 &0.020 \\ \hline
\end{tabular}}%
\end{center}}
In the above fusion process, we assume that different models have the same fusion weight, ignoring the reliability of each SVM model. Taking into account the impact of different fusion weights on the results, this paper uses 8\% of the dataset to conduct experiments. The results are shown in the Table 7.\ The first row in the table represents different fusion weights. For example, 0.6 indicates that the weight of the SVM model trained with Fre-Mea features is 0.6, and the weight of the SVM model trained with Fre-Var features is 0.4. According to the results, it can be concluded that as the weight of the SVM model trained by Fre-Mea increases, the accuracy of the proposed method increases. It is reasonable since the performance of the SVM model trained based on Fre-Var is much lower than its counterpart based on Fre-Mea. Nevertheless, the proposed fusion method is robust and not affected by the inferior SVM model too much. For example, if we set the weight to 0.9 and 0.1, the accuracy is only 0.02 less than the accuracy of the best single model, but its standard deviation is the lowest.
{\tabcolsep=2.5pt \footnotesize
\begin{center}
\scalebox{0.9}{
\begin{tabular}{|p{11.01em}<{\centering}|c|c|c|c|c|c|c|}
 \multicolumn{8}{c}{\bf Table 7.\ Test accuracy with different fusion weight}\\ \hline
    Fusion weight & 0 & 0.5   & 0.6  & 0.7  & 0.8 & 0.9 &1\\  \hline
    SVM-Fusion(Ave.) &0.759 & 0.956 & 0.951 & 0.944  &0.944 &0.958 &0.954 \\ \hline
    SVM-Fusion(Std.)  & 0.038 &0.019 & 0.019 & 0.019 &0.020 &0.018 &0.026 \\ \hline
    Proposed method(Ave.) & 0.82  & 0.974   & 0.975   & 0.977 &0.979  &0.982  &0.984 \\ \hline
    Proposed method(Std.) & 0.04 & 0.02 & 0.018 & 0.012 &0.015 &0.012 &0.013 \\  \hline
\end{tabular}}%
\end{center}}
In all the experiments, we used the ground truth labels which are clean. However, real labels always contain noise due to the ambiguity of the data or uncertainty of experts. In order to study the influence of label noise on the proposed framework, the datasets with 8\% volumes is adopted. First, we randomly selected 1/8 of the samples in the above mentioned scarce labeled dataset and assigned random category labels to them. Then, we conducted the experiments following the same procedure as in  Table 8.\ The results of the experiment on noisy labels are shown in the  Table 8.\ It can be seen that the noisy labels has an obvious side effect on the vanilla deep CNN model. Its classification accuracy dropped significantly from 97\% to 92.6\%. But the proposed method tends out to be more robust to label noise. Its classification only drops by 1.5\%, which is a much smaller margin compared with the counterpart of the vanilla deep CNN model.
{\tabcolsep=10.5pt \footnotesize
\begin{center}
\scalebox{0.85}{
\begin{tabular}{|p{10.65em}<{\centering}|c|c|c|c|}
\multicolumn{5}{c}{\bf Table 8.\ Classification results on the dataset with noisy labels}\\ \hline
      \multicolumn{1}{|c|}{\multirow{2}[2]{*}{}} & \multicolumn{2}{c|}{Clean labels} & \multicolumn{2}{c|}{Noisy labels} \\
    \cline{2-5}  \multicolumn{1}{|p{10.65em}<{\centering}|}{} & Ave. & Std. & Ave. & Std.\\ \hline
    CNN(Train) & 1     & -     & 1     & - \\  \hline
    CNN(Test) & 0.97  & 0.02  & 0.926 & 0.02 \\  \hline
    CNN-T(Fre-Mea;Train) & 1     & -     & 1     & - \\  \hline
    CNN-T(Fre-Mea;Test) & 0.984 & 0.015 & 0.965 & 0.015 \\  \hline
    CNN-T-Fusion(Train) & 1     & -     & 1     & - \\  \hline
    CNN-T-Fusion(Test) & 0.974 & 0.013 & 0.957 & 0.013 \\  \hline
\end{tabular}}%
\end{center}}

\
\begin{center}{\large\bf V.\ Conclusion and Discussions
}\end{center}

In this paper, we introduced a novel deep fault diagnosis (DFD) method for rotating machinery with scarce labeled samples. DFD successfully tackles the challenging problem by transferring knowledge from shallow models. DFD consists of three phases: 1) Selecting discriminative hand-crafted features to form the feature candidate pool, 2) Pre-training shallow models by using features from the feature candidate pool with scarce labeled samples. The best-performed shallow models are then selected based on the validation accuracy, which are used to make predictions on the unlabeled samples. They are combined together with the scarce fine labeled samples to form an augmented training set (ATS). 3) Knowledge-transferring from the shallow models to the proposed Deep CNN model by training it on the ATS. Experimental results on two fault diagnosis datasets demonstrate the effectiveness of the proposed DFD. Compared to the shallow model and the vanilla deep CNN model trained on scarce labeled samples, DFD achieves better performance. Moreover, it is computationally efficient, which is promising for real-time rotating machinery fault diagnosis.

The future work could include: 1) Exploiting other types of shallow models to compare them with SVM and clarify their correlations and complementarities, which further guides the knowledge transferring from ``specific experts''. 2) Extending the proposed DFD to time-varying system where online updating of the candidate feature pool and candidate model pool as well as incremental updating of the deep CNN are necessary.

\RE

\footnotesize\rm

\REF{[1]} W.J.\ Sun, S.Y.\ Shao, R.\ Zhao, {\it et al.}, ``A sparse auto
encoder-based deep neural network approach for induction motor faults
classification'', {\it Measurement}, Vol.89,
pp.171--178, 2016.

\REF{[2]} T.\ Ince, S.\ Kiranyaz, L.\ Eren, {\it et al.}, ``Real-time motor fault detection by 1D convolutional neural networks'',
{\it IEEE Transactions on Industrial Electronics}, Vol.63, No.11, pp.7067--7075, 2016.

\REF{[3]} M.\ Gan, C.\ Wang and C.A.\ Zhu,  ``Construction of hierarchical diagnosis
network based on deep learning and its application in the fault pattern recognition of rolling element bearings'',
 {\it Mechanical Systems and Signal Processing}, Vol.72--73, pp.92--104, 2016.

\REF{[4]} D.Z.\ Li, W.\ Wang and F.\ Ismailm, ``An enhanced bispectrum technique with auxiliary
frequency injection for induction motor health condition monitoring'', {\it IEEE Transactions on
Instrumentation \& Measurement}, Vol.64, No.10, pp.2679--2687, 2015.

\REF{[5]} S.\ Dash and V.\ Venkatasubramanian, ``Challenges in the industrial applications of
 fault diagnostic systems'',{\it Computers and Chemical Engineering},
Vol.24, No.2--7, pp.785--791, 2000.

\REF{[6]} A.\ Widodo and B.S.\ Yang, ``Support vector machine in machine
condition monitoring and fault diagnosis'',{\it Mechanical Systems and Signal Processing},
Vol.21, No.6, pp.2560--2574, 2007.

\REF{[7]} Y.G.\ Lei, Z.J.\ He and Y.Y.\ Zi, ``A new approach to intelligent fault
diagnosis of rotating machinery'',{\it Expert Systems with Applications}, Vol.35, No.4, pp.1593--1600, 2008.

\REF{[8]} J.\ Zhang, D.Q.\ Zhang, M.Y.\ Yang, {\it et al.}, ``Fault diagnosis for rotating machinery
with scarce labeled samples: a Deep CNN method based on knowledge-transferring from shallow models'', {\it International
Conference on Control, Automation and Information Sciences}, Hangzhou, China, pp.482--487, 2018.

\REF{[9]} W.\ Zhou, T.G.\ Habetler and R.G.\ Harley,  `` Bearing fault detection via stator
current noise cancellation and statistical control'',{\it IEEE Transactions on Industrial Electronics},
Vol.55, No.12, pp.4260--4269, 2008.

\REF{[10]} B.\ Sreejith, A.K.\ Verma and A.\ Srividya,  `` Fault diagnosis of rolling element bearing
using time-domain features and neural networks'',{\it IEEE Region 10 and the Third International Conference on Industrial and Information Systems},
Peradeniya,Sri Lanka, pp.1--6,2009.

\REF{[11]} Y.\ Liu, J.H.\ Zhang, K.J.\ Qin, {\it et al.}, ``Diesel engine fault diagnosis using intrinsic
time-scale decomposition and multistage Adaboost relevance vector machine'',
{\it Proceedings of the Institution of Mechanical Engineers, Part C: Journal of Mechanical Engineering Science}, Vol.232,
No.5, pp.881--894, 2018.

\REF{[12]} P.K.\ Wong, Z.X.\ Yang, C.M\ Vong, {\it et al.}, ``Real-time fault diagnosis for gas
turbine generator systems using extreme learning machin'', {\it Neurocomputing}, Vol.128, pp.249--257, 2014.

\REF{[13]} Z.Q.\ Chen, C.\ Li and R.V.\ Sanchez, ``Gearbox fault identification and classification
with convolutional neural networks'',{\it Shock and Vibration}, Vol.2015, Article ID 390134, 10 pages, 2015.

\REF{[14]} Y.\ Lv, R.\ Yuan and G.B.\ Song,  ``Multivariate empirical mode decomposition and its
application to fault diagnosis of rolling bearing'',
 {\it Mechanical Systems and Signal Processing}, Vol.81, pp.219--234, 2016.

\REF{[15]} J.D.\ Zheng, H.Y.\ Pan, X.L\ Qi, {\it et al.}, ``Enhanced empirical wavelet
transform based time-frequency analysis and its application
to Rolling Bearing Fault Diagnosis'', {\it Acta Electronica Sinica}, Vol.46,
No.2, pp.358--364, 2018.(in Chinese)

\REF{[16]} C.\ Li, V.\ Sanchez, G.\ Zurita, {\it et al.}, ``Rolling element bearing
defect detection using the generalized synchrosqueezing transform guided by
time¨Cfrequency ridge enhancement'', {\it Isa Transactions}, Vol.60, pp.274--284, 2016.

\REF{[17]} Y.\ Tian, J.\ Ma, C.\ Lu, {\it et al.}, ``Rolling bearing fault diagnosis
under variable conditions using LMD-SVD and extreme learning machine'', {\it Mechanism and Machine Theory},
Vol.90, pp.175--186, 2015.

\REF{[18]} M.A.\ Hearst, S.T.\ Dumais, E.\ Osman, {\it et al.}, ``Support vector machines'',
{\it IEEE Intelligent Systems},
Vol.13, No.4, pp.18--28, 1998.

\REF{[19]} X.Y.\ Zhang, Y.T.\ Liang, J.Z.\ Zhou, {\it et al.}, ``A novel bearing fault diagnosis model
integrated permutation entropy, ensemble empirical mode decomposition and optimized SVM'',
{\it Measurement}, Vol.69, pp.164--179, 2015.

\REF{[20]} A.N.\ Wang, M.\ Sha, L.M.\ Liu, {\it et al.}, ``A new process industry fault
diagnosis algorithm based on ensemble improved binary-tree SVM'',
{\it Chinese Journal of Electronics}, Vol.24, No.2, pp.258--262, 2015.

\REF{[21]} Y.\ Freund and R.E.\ Robert, ``A decision-theoretic generalization of on-line learning and
an application to boosting'',{\it Journal of computer and system sciences}, Vol.55, No.1, pp.119--139, 1997.

\REF{[22]} H.W.\ Liu, L.\ Liu and H.J.\ Zhang,  ``Boosting feature selection using
information metric for classification'',{\it Neurocomputing}, Vol.73, No.1--3,pp.295--303, 2009.

\REF{[23]} G.B.\ Huang, Q.Y.\ Zhu and C.K.\ Siew, ``Extreme learning machine: Theory and applications'',
{\it Neurocomputing}, Vol.70, No.1--3, pp.489--501, 2006.

\REF{[24]} G.E.\ Hinton, S.\ Osindero and Y.W.\ Teh, ``A fast learning algorithm for deep belief nets'',
{\it Neural computation}, Vol.18, No.7, pp.1527--1554, 2006.

\REF{[25]} H.D.\ Shao, H.K.\ Jiang, X.\ Zhang, {\it et al.}, ``Rolling bearing fault diagnosis using
an optimization deep belief network'',{\it Measurement Science and Technology}, Vol.26, No.11, 2015.

\REF{[26]} Y.\ Lecun and Y.\ Bengio, ``Convolutional networks for images, speech, and time series'',
{\it The handbook of brain theory and neural networks},1995.

\REF{[27]} F.\ Gao, M.\ Wang, J.\ Wang,{\it et al.,}  ``A novel separability objective function
in CNN for feature extraction of SAR images'',{\it Chinese Journal of Electronics},  Vol.28, No.2, pp.423--429,2019.

\REF{[28]} D.D.\ Bai, C.Q.\ Wang, B.\ Zhang, {\it et al.}, ``CNN feature boosted seqSLAM for real-Time
loop closure detection'',{\it Chinese Journal of Electronics}, Vol.27, No.3, pp.488--499, 2018.

\REF{[29]} J.Y.\ Gan, Y.K.\ Zhai, Y.\ Huang, {\it et al.}, ``Research of facial beauty prediction based on
deep convolutional features using double activation layer'',{\it Acta Electronica Sinica},
Vol.47, No.3, pp.636--642, 2019.(in Chinese)

\REF{[30]} L.\ Deng, J.Y\ Li J.T.\ Huang, {\it et al.}, ``Recent advances in deep learning
for speech research at Microsoft'', {\it IEEE International Conference on Acoustics}, Vancouver, British Columbia, Canada,
 pp.8604--8608,2013.

\REF{[31]} A.\ Krizhevsky, I.\ Sutskever and G.E.\ Hinton, ``ImageNet Classification with Deep
Convolutional Neural Networks'', {\it International Conference on Neural Information
Processing Systems}, Doha, Qatar, pp.1097--1105,2012.

\REF{[32]} S.Q.\ Ren, K.M.\ He, R.\ Girshick, {\it et al.}, ``Faster R-CNN: towards real-time
object detection with region proposal networks'', {\it International Conference on Neural Information
Processing Systems}, Kuching, Malaysia, pp.91--99,2015.

\REF{[33]} F.\ Jia, Y.G.\ Lei, J.\ Lin, {\it et al.}, ``Deep neural networks: A promising tool for fault
characteristic mining and intelligent diagnosis of rotating machinery with massive data'',
{\it Mechanical Systems and Signal Processing}, Vol.72--73, pp.303--315, 2016.

\REF{[34]} M.\ Xia, T.\ Li, L.\ Xu, {\it et al.}, ``Fault diagnosis for rotating machinery
using multiple sensors and convolutional neural networks'',
{\it IEEE/ASME Transactions on Mechatronics}, Vol.23, No.1, pp.101--110, 2017.

\REF{[35]} J.H.\ Sun, Z.W.\ Xiao and Y.X.\ Xie, ``Automatic multi-fault recognition
in TFDS based on convolutional neural network'',{\it Neurocomputing}, Vol.222, pp.127--136, 2017.

\REF{[36]} M.\ Meng, Y.J.\ Chua, E.\ Wouterson, {\it et al.}, ``Ultrasonic signal classification and
imaging system for composite materials via deep convolutional neural networks'',
{\it Neurocomputing}, Vol.257, pp.128--135, 2017.

\REF{[37]} L.\ Deng, M.L.\ Seltzer, D.\ Yu, {\it et al.}, ``Binary coding of speech spectrograms using a deep auto-encoder'',
{\it 11th Annual Conference of the International Speech Communication Association},Makuhari,Japan,2010.

\REF{[38]} J.M.\ Zhang, Q.Q.\ Huang, H.L.\ Wu, {\it et al.}, ``A shallow network with combined
pooling for fast traffic sign recognition'',{\it Information}, Vol.8, No.2, pp.45, 2017.

\REF{[39]} S.\ Santurkar, D.\ Tsipras, A.\ Ilyas, {\it et al.}, ``How does batch normalization help optimization?'',
{\it Advances in Neural Information Processing Systems}, Montreal,Canada, pp.2483--2493, 2018.

\REF{[40]} Y.Q.\ Jia, E.\ Shelhamer, J.\ Donahue, {\it et al.}, ``Caffe: Convolutional architecture for fast
feature embedding'', {\it Proc.of the 22nd ACM international conference on Multimedia}, Orlando,Florida USA, pp.675--678, 2014.

\REF{[41]} L.V.D.\ Maaten and G.\ Hinton, ``Visualizing data using t-SNE'',
{\it Journal of machine learning research}, Vol.9, No.Nov, pp.2579--2605, 2008.

\REF{[42]} F.N.\ Zhou, P.\ Hu, S.\ Yang, {\it et al.}, ``A multimodal feature fusion-based
deep learning method for online fault diagnosis of rotating machinery'',
{\it Sensors}, Vol.18, No.10, pp.3521, 2018.

\REF{[43]} H.Y.\ Pan, J.D.\ Zheng, Y.\ Yang, {\it et al.}, ``Research on combined intelligent
fault diagnostic method based on CELCD and MFVPMCD'',
{\it Acta Electronica Sinica}, Vol.45, No.3, pp.546--551, 2017.(in Chinese)





\end{multicols} 
\end{document}